\documentclass[12pt,3p,authoryear]{elsarticle}
\usepackage[T1]{fontenc}
\usepackage{amsmath,amssymb,amsfonts}
\usepackage{natbib}
\usepackage{graphicx}
\usepackage{epstopdf}
\usepackage[hidelinks,breaklinks=true]{hyperref}
\usepackage{booktabs}
\usepackage{rotating}
\usepackage{times}
\usepackage{xspace}
\usepackage{dcolumn}
\newcolumntype{.}[1]{D{.}{.}{#1}}
\newcolumntype{,}[1]{D{,}{,}{#1}}
\usepackage{setspace}
\usepackage{color}
\usepackage{subdepth}
\usepackage{needspace}   % almost always useful
\usepackage{float}
\usepackage[font={small,singlespacing},labelfont={bf,small},skip=0.2cm]{caption}
\floatstyle{plaintop}
\restylefloat{table}
\restylefloat{figure}			
\journal{Finance Research Letters -- Special Issue on Carbon Finance}
\graphicspath{{./img/}}
\usepackage{multirow}
\RequirePackage{ae,fancyvrb}
\newcommand{\pkg}[1]{\textbf{#1}}

\newcommand{\proglang}[1]{\textsf{#1}}
\usepackage{upquote}
\DefineVerbatimEnvironment{CodeInput}{Verbatim}{fontshape=sl}
\DeclareMathAlphabet\mathbfcal{OMS}{cmsy}{b}{n}

\usepackage[font={normal}]{subfig}

\usepackage{color}

\newcommand{\ie}{\emph{i.e.}\xspace}
\newcommand{\eg}{\emph{e.g.}\xspace}

\newcommand{\insertfloat}[1]{%
\begin{center}
[Insert~#1 about here.]%
\end{center}%
}

\begin{document}
\begin{frontmatter}
\title{How easy is it for investment managers to deploy their talent in green and brown stocks?\tnoteref{label1}}
\tnotetext[label1]{We thank Marco Kerkemeier, Yves Robinson Kruse-Becher, and participants at the Workshop on Carbon Finance 2022 for their comments. We are grateful to IVADO, NSERC, and the Swiss National Science Foundation (grant \#179281) for their financial support.}
\author[hec]{David Ardia\corref{cor1}}
\ead{david.ardia@hec.ca}
\cortext[cor1]{Corresponding author. HEC Montréal, 3000 Chemin de la Côte-Sainte-Catherine, Montreal, QC H3T 2A7. Phone: +1 514 340 6103.}
\ead{keven.bluteau@usherbrooke.ca}
\author[sherbrooke]{Keven Bluteau}
\ead{keven.bluteau@usherbrooke.ca}
\author[hec]{Thien Duy Tran}
\ead{thien-duy.tran@hec.ca}
\address[hec]{GERAD \& Department of Decision Sciences, HEC Montréal, Montréal, Canada}
\address[sherbrooke]{Department of Finance, Université de Sherbrooke, Canada\\[1cm]
\large Published in Finance Research Letters\\
\url{https://doi.org/10.1016/j.frl.2022.102992}}

\begin{abstract}
We explore the realized alpha-performance heterogeneity in green and brown stocks' universes using the peer performance ratios of \citet{ArdiaBoudt2018}. Focusing on S\&P~500 index firms over 2014--2020 and defining peer groups in terms of firms' greenhouse gas emission levels, we find that, on average, about 20\% of the stocks differentiate themselves from their peers in terms of future performance. We see a much higher time-variation in this opportunity set within brown stocks. Furthermore, the performance heterogeneity has decreased over time, especially for green stocks, implying that it is now more difficult for investment managers to deploy their skills when choosing among low-GHG intensity stocks.
\end{abstract}
\end{frontmatter}

\onehalfspacing
%\singlespacing

\newpage
\section{Introduction}

Environmental, social, and governance (ESG)-focused investing has become a hot topic in recent years. According to Morningstar, global sustainable fund assets hit a record US\$3.9 trillion in 2021-Q3.\footnote{See \url{https://www.reuters.com/business/sustainable-business/global-sustainable-fund-assets-hit-record-39-trillion-q3-says-morningstar-2021-10-29/}} Among the ESG dimensions, the environmental concern is playing a leading role. As indicated by \citet{PastorEtAl2021}, 88\% of the clients of BlackRock, the world's largest asset manager, rank environment as ``the priority most in focus'' among ESG criteria. Institutional investors and asset managers are increasingly tracking the greenhouse gas emissions of listed firms when building their portfolios \citep{KruegerEtAl2020,BoltonKacperczyk2021}. 

In light of this, investment managers are increasingly subject to constraints regarding which assets they can invest in. Our research note aims to investigate to what extent they can differentiate themselves in terms of future performance when focusing on green or brown stocks' universes. We posit that the more heterogeneous a universe is, in terms of its underlying stocks' performance, the more easily good (bad) managers can deploy their skills (unskills) and differentiate themselves from their peers. 

To explore the performance heterogeneity in green and brown stocks' universes, we rely on the peer performance ratios by \citet{ArdiaBoudt2018}. The output of their methodology is a triplet: (i) an equal-performance ratio, (ii) an underperformance ratio, and (iii) an outperformance ratio. The former measures the percentage of stocks in a given peer group that cannot be differentiated in terms of performance. The two other ratios attribute the remaining portion to over- or underperforming stocks. The approach is an extension to the false discovery rate approach by  \citet{Storey2002}, pioneered by \citet{BarrasEtAl2010} in their analysis of mutual funds performance; see \citet{ArdiaBoudt2018} for an application to hedge funds. We refer to  \citet{harveyEtAl2020b} for a review on multiple testing methods that have been developed to control for luck in active management.

Our empirical study focuses on  S\&P 500 index firms for 2014--2020. We use firms' greenhouse gas (GHG) emission intensity to create peer groups of green and brown stocks.\footnote{ \citet{GibsonEtAl2021} find that the correlation between seven ESG-data providers is the highest for the environmental score (average correlation at 0.47), and rationalize this by the fact that 
greenhouse gas emissions are a key dimension of a firm's environmental performance.}
 Using \citet{Carhart1997} and \citet{FamaFrench2015} factor models, we find that, on average, about 20\% of the stocks differentiate themselves from their peers in terms of realized-alpha performance over various horizons (from three months to one year of daily data). We see a much higher variability in this opportunity set within brown stocks. Furthermore, this heterogeneity has decreased over time, especially for green stocks, implying that it is now more difficult for investment managers to deploy their skills when allocating among low-GHG intensity stocks.

\section{Data}
\label{sec:data}

Our study focuses on firms in the S\&P 500 index from January 2014 to December 2020. We define green (brown) firms as firms that create economic value while minimizing (not minimizing) damages that contribute to climate change. We follow \citet{ArdiaEtAl2020} and use CO2-equivalent emissions scaled by the firm's revenues, namely, GHG emission intensity. We define total GHG emissions as the sum of scope 1 (direct CO2 emissions), scope 2 (indirect CO2 emissions), and scope 3 (indirect CO2 emissions from the value chain) in concordance with the Greenhouse Gas Protocol Standards.\footnote{See \url{https://ghgprotocol.org/standards}.} CO2-equivalent emissions data come from Thomson/Refinitiv, and firms' revenues are from Compustat. Thus, our key ranking variable is the total tons of CO2-equivalent GHG emissions attributed to a one million dollar revenue, which we will refer simply to as GHG or GHG emissions intensity. 

A summary of our data set is reported in Table \ref{tab:data}. Since 2014, more than 60\% of the S\&P 500 firms have disclosed their CO2 emissions data. The distribution of GHG emissions intensity is highly positively skewed and fat-tailed. 

\insertfloat{Table~\ref{tab:data}}

Finally, daily stock prices data come from CRSP, and daily factor data are retrieved from Kenneth French's website.\footnote{See \url{http://mba.tuck.dartmouth.edu/pages/faculty/ken.french/data_library.html}.}

\section{Methodology}
\label{sec:methodology}

To determine to what extent investment managers could differentiate themselves when investing in green or brown stocks in 2014--2020, we proceed as follows:

\paragraph{Step 1} In a given month, we form peer groups of brown and green stocks using the information available up to that month. We rely on the firms' latest GHG emissions released to form the groups.\footnote{GHG data are usually released with a one-year delay.} Brown (green) stocks belong to the top 75th (bottom 25th) percentile of GHG intensities.\footnote{We tested alternative thresholds, and results were qualitatively similar. All alternative setups tested in the paper are available from the autors upon request.} Each group contains the same number of peers. 

\paragraph{Step 2} For each firm $i=1,\ldots,N$ within a peer group, we compute the three peer-performance ratios, $\pi_i^0$,  $\pi_i^-$, $\pi_i^+$, following  \citet{ArdiaBoudt2018}. Expanding on \citet{BarrasEtAl2010}, \citet{ArdiaBoudt2018} develop a triple-layered peer performance evaluation framework. For them, a stock can exhibit three types of peer performance with respect to a peer group:\footnote{\citet{ArdiaBoudt2018} apply the methodology to hedge funds, but the approach can be applied to any set of instruments, such as stock returns as we do here.}
\begin{enumerate}
\item[(i)]  equal-performance ($\pi_i^0$): the percentage of peers stocks that perform equally as stock~$i$;
\item[(ii)] underperformance ($\pi_i^-$): the percentage of peer stocks that outperform stock~$i$;
\item[(iii)] outperformance ($\pi_i^+$): the percentage of peer stocks that underperform stock~$i$.
\end{enumerate}
For a given stock $i$, the peer performance ratios are obtained with a two-step procedure \citep[see][Section 2]{ArdiaBoudt2018}:

\paragraph{Step 2a} For each of the $N-1$ peer stocks $\mathcal{J}_i = \{j=1,\ldots,N; j \neq i\}$, we calculate the $p$-value of the null hypothesis of equal-performance over a forward-looking evaluation period between stock $i$ and stock $j \in \mathcal{J}_i$ using a pairwise test. We consider a $h$-month-ahead horizon of $T$ daily returns and estimate the alpha differential from a factor model:
\begin{equation}
r_{i,t}-r_{j,t}=\Delta \alpha_{i,j}+\sum_{k=1}^K \beta_{i,j,k} F_{k,t} + \varepsilon_{i,j,t} \,\, (j \in \mathcal{J}_i \,;\, t=1,\ldots,T) \,,
\end{equation}
where $r_{i,t}$ is the daily return of stock $i$ at time $t$, $F_{k,t}$ is the daily return of the $k$th factor at time $t$, and $\varepsilon_{i,j,t}$ is an error term. In our empirical application, we use the four-factor model by \citet{Carhart1997} and the
five-factor model by \citet{FamaFrench2015}. The four-factor model encompasses the market (MKT), the small-minus-big (SMB), the high-minus-low (HML), and the momentum factors. The five-factor model encompasses the MKT, the SMB, the HML, the robust-minus-weak (RMW), and conservative-minus-aggressive (CMA) factors. For a given factor model, this leads to a set of coefficients $\{\Delta \hat \alpha_{i,j}, j \in \mathcal{J}_i\}$. Each coefficient is then standardized using the heteroscedastic and autocorrelation
robust (HAC) standard error estimator of \citet{Andrews1991} and \citet{AndrewsMonahan1992}, leading to a studentized test statistic from which the $p$-value is obtained via the probability integral transform.

\paragraph{Step 2b} The distribution of the $p$-values obtained in \emph{Step 2a} is (asymptotically) a mixture of $p$-values that are uniformly distributed when the null hypothesis is true and $p$-values close to zero when the null hypothesis is false.\footnote{\citet{BarrasEtAl2020} and \citet{AndrikogiannopoulouPapakonstantinou2019} recommend careful evaluation of the underlying data generating process when using the false discovery rate approach with financial data, especially when the sample size is small. This is not the case here, as we are dealing with a minimum of 60 observations (\ie, three months of daily returns). Moreover, as noted in \citet[Footnote 8]{ArdiaBoudt2018}, a very high correlation between the test statistics may lead to an inconsistent estimator. We checked this in our sample and found correlations in reasonable ranges (\ie, within [-0.5,0.5] in the vast majority of the test statistics).} Following \citet[Section 2.4]{ArdiaBoudt2018}, this allows us to set a data-driven cut-off point to estimate the proportion of equal-performance $\pi_i^0$ which is robust to false discoveries. The remaining proportion is attributed to the outperformance and underperformance ratios \citep[Section 2.5]{ArdiaBoudt2018}.\footnote{All of the computations employed the \proglang{R} statistical computing language \citep{R} with the package \textbf{PeerPerformance} \citep{PeerPerformance}, which is freely available at \url{https://CRAN.R-project. org/package=PeerPerformance}.}

\paragraph{Step 3} Once the ratios are computed for the $N$ firms within a peer group, we compute the aggregate ratios. First, $\pi^0 = \frac{1}{N} \sum_{i=1}^N \pi_i^0$ measures the equal-performance ratio within a peer group. As such, $1-\pi^0$ can be used to assess the extent to which stocks out- or underperform compared to their peers in the group: it is a measure of performance heterogeneity. The higher it is, the more direct it is for investment managers to deploy their talent at either (i) picking outperforming stocks or (ii) selling (or shorting) underperforming stocks in the peer group of brown or green funds to differentiate themselves from their competitors. Second, the quantities $\pi^- =  \frac{1}{N} \sum_{i=1}^N \pi_i^-$ and $\pi^+ =  \frac{1}{N} \sum_{i=1}^N \pi_i^+$ (which sum to $(1-\pi^0)$ by construction) are then used to assess the part of underperforming or outperforming stocks in the peer group.

Steps 1--3 are repeated every month starting from January 2014, yielding a time series of aggregate peer-performance ratios. The ending date depends on the horizon $h$ considered: September 2020 for $h=3$, June 2020 for $h=6$, and December 2019 for $h=12$ months. Note that, by construction, the monthly ratios are autocorrelated, as they are based on overlapping daily returns (\eg, in the $h=12$ months horizon, 11 months of daily returns are overlapping between two consecutive months in the time series of the peer performance ratios).

\section{Results}
\label{sec:ratios}

Following the steps in Section~\ref{sec:methodology}, we estimate the performance heterogeneity and the under- and outperformance ratios for the universes of green and brown stocks monthly over 2014--2020. Results are reported in Table~\ref{tab:ratios}. 

First, we find that the unconditional performance heterogeneity is around 20\%. The percentage is systematically higher for brown stocks than for green stocks. Results are robust over the three evaluation horizons (Columns 2--3, 4--5, and 6--7) and the two factors models (Panels A and B). Second, we observe a much higher variability in performance heterogeneity for brown stocks as reported by the standard deviation and the maximum-minimum range. For instance, at the three-month horizon with the four-factor model, the maximum performance heterogeneity percentage in the brown stocks group is 48.9\%, while 29.8\% for green stocks. This large difference is also observed with the five-factor model. Hence, the range of opportunity has been slightly higher for the brown stocks in 2014--2020 and has exhibited periods of higher heterogeneity for the set of brown stocks than for green stocks. In terms of investment opportunities, this means that it has been easier for investment managers to deploy their talent in brown stocks than green stocks. Finally, we see a negative trend in the performance heterogeneity (\ie, the slope of a linear regression with time as the explanatory variable) for green and brown firms. The negative trend is significant at the one percent level for green stocks at the twelve-month horizon.\footnote{Statistical significance is based on heteroscedasticity and autocorrelation robust standard error estimators \citep{Andrews1991,AndrewsMonahan1992}. Conclusions hold if we rely on Hansen-Hodrick standard errors \citep{HansenHodrick1980}, stationary-bootstrap standard errors \citep{PolitisRomano1994,PolitisWhite2004}, or fixed-length block-bootstrap standard errors for various block lenghts. Alternative approaches that could be considered for testing the trend are \citet{GadeaRivas2020} and \citet{BunzelVogelsang2005}, for instance.} Results are robust to the inclusion of the performance heterogeneity of neutral firms (\ie, firms with GHG intensity within the 25th--75th percentile range) as a control in the linear regression or when using a beta regression to measure the trend \citep{FerrariCribariNeto2004}.\footnote{The difference in time-trends between brown and green stocks' universes is significantly different from zero at the 5\% level for the five-factor model over the 12-month evaluation period. Here again, results are robust to the choice of the standard-error estimators.}   

When looking at the components of the performance heterogeneity, that is, the underperformance and outperformance ratios in Panels C and D, we see that it is primarily the underperformance ratios that drive the heterogeneity performance of brown stocks. For instance, in the case of the three-month horizon, the average is 11.6\% for~$\pi^-$ and 9.4\% for~$\pi^+$. It is also the case for green stocks, but to a lesser extent. We also see that the largest variability of both ratios is observed for the underperformance ratio. For both peer groups, the trend is negative but only significant for all trend specifications in the case of the outperformance ratio of green stocks over a twelve-month evaluation horizon. Overall, whether trading in the brown or green space, investment managers would have been able to differentiate themselves more easily by avoiding (or shorting) bad performer stocks. Moreover, picking outperforming green stocks has become
more challenging in recent years.

\insertfloat{Table~\ref{tab:ratios}}

\begin{figure}[H]
\caption{\textbf{Performance heterogeneity over time}\\
The following graphs present the evolution of performance heterogeneity $1-\pi^0$ over time for the brown stocks' universe (left panel) and green stocks' universe (right panel). The performance heterogeneity is computed with three-month daily returns (top plots) and with twelve-month daily returns (bottom plots). Both are obtained with the four-factor model by \citet{Carhart1997}. The dashed line displays the time trend.}
\label{fig:ratios}
\centering
\includegraphics[scale=.28]{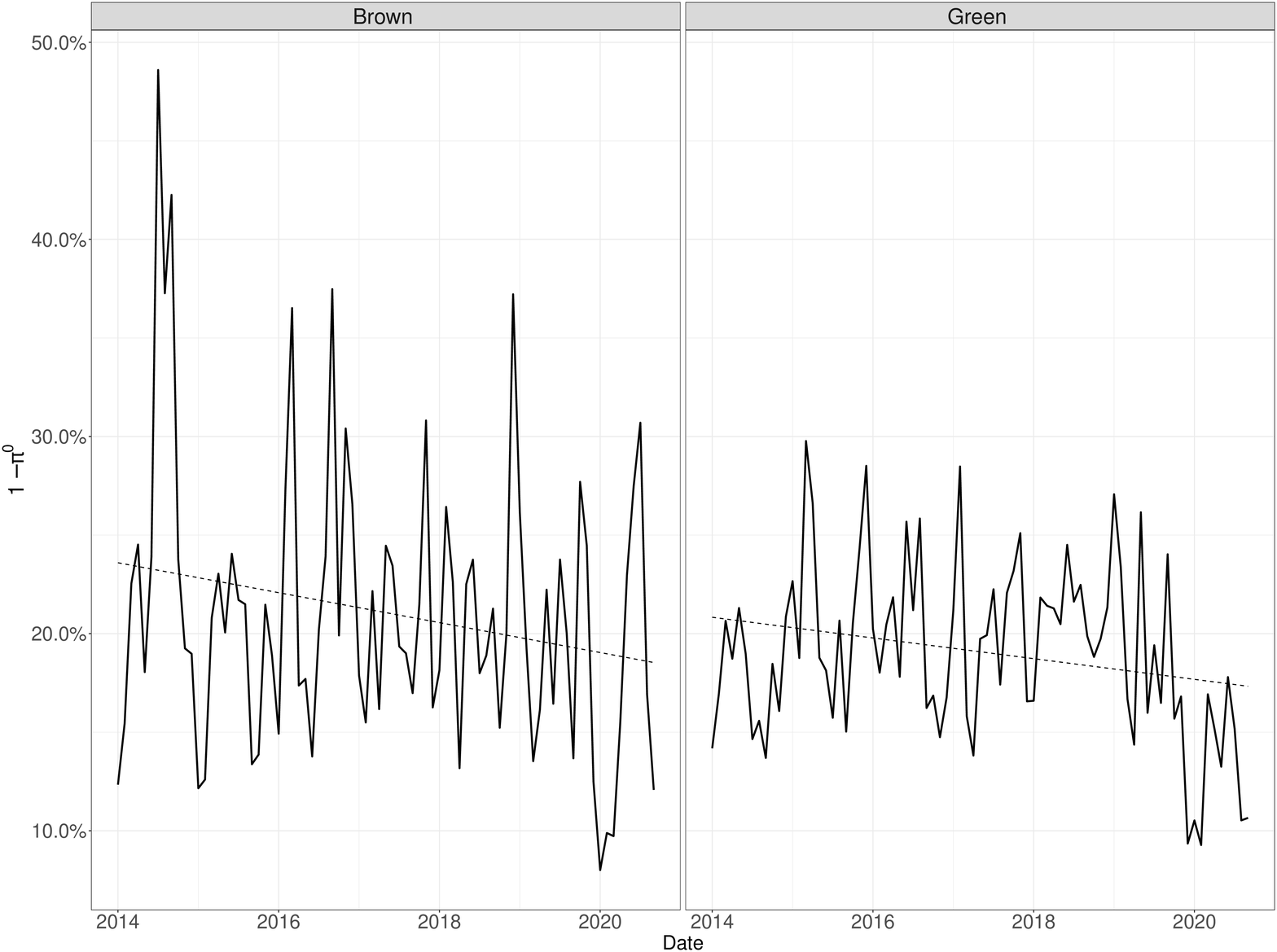}\\[.5cm]
\centering
\includegraphics[scale=.28]{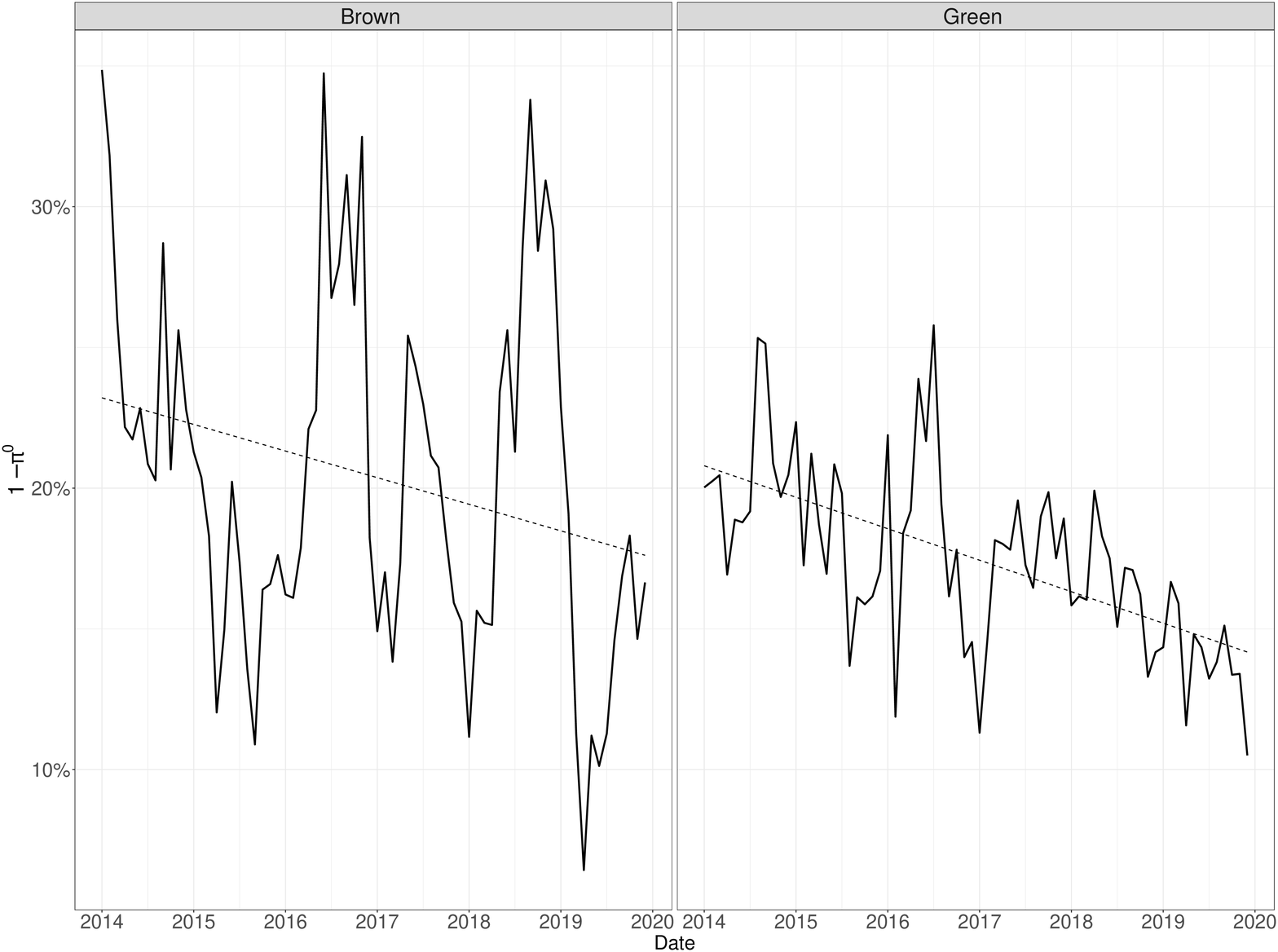}
\end{figure}

In Figure~\ref{fig:ratios}, we display the time series of the performance heterogeneity obtained with the four-factor model setup over a three-month (top panels) or twelve-month (bottom panels)  evaluation period.\footnote{Results for the five-factor model lead to the same conclusion.} We notice the declining percentage of performance heterogeneity for both brown and green stocks, as indicated by the trend in the dashed line. The higher variability is also evident for the brown stocks' universe. It is also interesting to note the dip in early 2020 for both universes in the three-month figure. The market crash in the first quarter of 2020 due to the COVID-19 pandemic drove the performance heterogeneity to its lowest value in green and brown universes. Post-crash, we see that the performance heterogeneity in green stocks remains particularly low compared to its historical average and brown stocks. Overall, while sustainable investing has become very trendy since the crisis, our findings suggest it is more challenging nowadays for investment managers in this space to differentiate themselves from their peers.

\insertfloat{Figure~\ref{fig:ratios}}

\newpage
\begin{table}[H]
\singlespacing
\caption{\textbf{Summary statistics of GHG}\\
This table reports the summary statistics of the greenhouse gas emissions level scaled by firms' revenues (GHG emissions intensity). Total GHG emissions level is defined as the sum of scope 1 (direct CO2 emissions), scope 2 (indirect CO2 emissions), and scope 3 (indirect CO2 emissions from the value chain) in concordance with the Greenhouse Gas Protocol Standards. CO2 emissions are from Thomson/Refinitiv, and firms' revenues are from Compustat.}
\centering
\scalebox{1.0}{
\begin{tabular}{lr}
\toprule
\multicolumn{2}{l}{Panel A: Percentage of S\&P 500 firms with CO2 emissions data} \\
Year & Value\\ 
\midrule
2014  & 61.7 \\ 
2015  & 62.2 \\ 
2016  & 66.4 \\ 
2017  & 70.2 \\ 
2018 & 73.8 \\
2019 & 80.2 \\
2020 & 78.1 \\
\midrule
\multicolumn{2}{l}{Panel B: Summary statistics of the GHG intensity data} \\ 
Statistics & Value\\ 
\midrule
Average                    &  811.6      \\ 
Standard  deviation       & 2,271.1   \\ 
Skewness                   & 4.4      \\ 
Excess kurtosis               &  22.2      \\ 
Minimum                    & 0.0       \\ 
25th   percentile          & 37.2      \\ 
50th   percentile         & 77.1      \\ 
75th   percentile         & 358.6     \\ 
Maximum                 & 18,561.1   \\ 
\bottomrule
\end{tabular}}
\label{tab:data}
\end{table}

\newpage
\begin{table}[H]
\singlespacing
\caption{\textbf{Summary statistics of the peer performance ratios}\\
\footnotesize This table reports the summary statistics of the performance heterogeneity $1 - \pi^0$, the underperformance ratio $\pi^-$, and the
outperformance ratio $\pi^+$. Panel A reports results for $1 - \pi^0$ based on the alpha of the four-factor model by \citet{Carhart1997}. Panel B reports the results for the five-factor model by \citet{FamaFrench2015}. Panel C (Panel D) reports the results for $\pi^-$ ($\pi^+$) for the four-factor 
model. ``Trend'' reports the monthly trend of the ratio obtained by linear regression. ``Trend with control'' considers the performance heterogeneity of neutral funds as a control variable in the regression. ``Trend with control (beta)'' is a beta regression with the performance heterogeneity of neutral funds as a control variable in the regression. All trend coefficients are multiplied by 1,000 for readability. The symbols $^{***}$, $^{**}$ and $^{*}$ indicate statistical significance of the trend at the 1\%, 5\%, and 10\% levels, respectively, based on heteroscedasticity and autocorrelation robust standard error estimators \citep{Andrews1991,AndrewsMonahan1992}. Three-month ahead refers to peer performance ratios computed with the three-month daily returns following the evaluation month. Six-month ahead and 
one-year ahead refer to six months and twelve months of daily returns, respectively. Brown (Green) refers to the peer group of firms in the top 75th percentile (bottom 25th percentile) of the GHG emissions at the evaluation date. Setting with $h=3$ months spans January 2014 to September 2020 (81 evaluation dates), $h=6$ months spans January 2014 to June 2020 (79 evaluation dates), and $h=12$ months spans January 2014 to December 2019 (72 evaluation dates).}
\centering
\vspace{-0.5cm}
\scalebox{0.78}{
\begin{tabular}{lrrrrrr}
\toprule
\multicolumn{7}{l}{Panel A:  Four-factor model used to compute $1 - \pi^0$} \\ 
&\multicolumn{2}{c}{$h = 3$ months} 
&\multicolumn{2}{c}{$h = 6$ months} 
&\multicolumn{2}{c}{$h = 12$ months} \\
\cmidrule(lr){2-3}\cmidrule(lr){4-5}\cmidrule(lr){6-7}
& Brown  & Green  & Brown  & Green  & Brown  & Green  \\ 
\midrule
Average &  21.1 & 19.1 & 20.0 & 18.1 & 20.4 & 17.5 \\ 
Standard  deviation   & 7.4 &  4.4 &  7.1 &  3.7 &  6.5 &  3.3 \\ 
Minimum &  8.0 &  9.3 &  6.4 & 10.0 &  6.4 & 10.5 \\ 
Maximum & 48.6 & 29.8 & 40.2 & 32.3 & 34.9 & 25.8 \\ 
Trend & -0.6 & -0.4 & -1.2$^{*}$  & -0.4 & -0.8 & -0.9$^{***}$  \\ 
Trend with control & -0.2 & -0.2 & -0.6 & -0.2 &  0.3 & -0.6$^{***}$  \\ 
Trend with control (beta regression) & -1.1 & -1.4 & -4.0  & -1.6 &  0.7 & -4.6$^{***}$  \\ 
\midrule
\multicolumn{7}{l}{Panel B:  Five-factor model used to compute $1 - \pi^0$} \\ 
&\multicolumn{2}{c}{$h = 3$ months} 
&\multicolumn{2}{c}{$h = 6$ months} 
&\multicolumn{2}{c}{$h = 12$ months} \\
\cmidrule(lr){2-3}\cmidrule(lr){4-5}\cmidrule(lr){6-7}
& Brown  & Green  & Brown  & Green  & Brown  & Green  \\ 
\midrule
Average & 20.5 & 18.7 & 18.9 & 17.4 & 17.3 & 16.8 \\ 
Standard  deviation   & 7.8 &  4.7 &  6.9 &  4.2 &  7.3 &  4.2 \\ 
Minimum & 3.5 &  7.2 &  5.1 &  4.9 &  4.3 &  7.4 \\ 
Maximum & 46.5 & 32.9 & 38.9 & 28.0 & 42.6 & 27.5 \\ 
Trend & -0.7 & -0.5 & -1.2$^{*}$ & -0.4 & -1.3 & -1.3$^{***}$ \\ 
Trend with control   & -0.3 & -0.2 & -0.6 &  0.0 &  0.4 & -0.7$^{**}$ \\ 
Trend with control (beta regression) & -2.2 & -1.8 & -4.0 & -0.3 &  2.4 & -5.5$^{**}$  \\ 
\midrule
\multicolumn{7}{l}{Panel C:  Four-factor model used to compute $\pi^-$} \\ 
&\multicolumn{2}{c}{$h = 3$ months} 
&\multicolumn{2}{c}{$h = 6$ months} 
&\multicolumn{2}{c}{$h = 12$ months} \\
\cmidrule(lr){2-3}\cmidrule(lr){4-5}\cmidrule(lr){6-7}
& Brown  & Green  & Brown  & Green  & Brown  & Green  \\ 
\midrule
Average & 11.6 &  9.6 & 11.0 &  9.4 & 11.4 &  9.2 \\ 
Standard  deviation  &  5.3 &  3.3 &  5.0 &  2.7 &  5.1 &  2.4 \\ 
Minimum &  1.0 &  4.1 &  1.5 &  3.9 &  3.0 &  3.9 \\ 
Maximum & 28.0 & 18.9 & 25.8 & 18.2 & 22.9 & 14.8 \\ 
Trend & -0.3 & -0.2 & -0.7 & -0.2 & -0.6 & -0.4$^{**}$ \\ 
Trend with control  & -0.1 & -0.1 & -0.5 & -0.1 & -0.1 & -0.2 \\ 
Trend with control (beta regression) &  -1.3 &  -1.6 &  -5.3 &  -1.1 &  -0.5 &  -2.2 \\ 
\midrule
\multicolumn{7}{l}{Panel D:  Four-factor model used to compute $\pi^+$} \\ 
&\multicolumn{2}{c}{$h = 3$ months} 
&\multicolumn{2}{c}{$h = 6$ months} 
&\multicolumn{2}{c}{$h = 12$ months} \\
\cmidrule(lr){2-3}\cmidrule(lr){4-5}\cmidrule(lr){6-7}
& Brown  & Green  & Brown  & Green  & Brown  & Green  \\ 
\midrule
Average &  9.4 &  9.5 &  9.0 &  8.7 &  9.0 &  8.3 \\ 
Standard  deviation  &  3.6 &  2.7 &  3.3 &  2.6 &  3.0 &  2.2 \\ 
Minimum &  3.6 &  4.1 &  4.2 &  2.7 &  3.0 &  3.8 \\ 
Maximum & 20.6 & 15.7 & 16.3 & 15.8 & 16.5 & 13.2 \\ 
Trend & -0.3 & -0.2 & -0.5$^{**}$ & -0.3 & -0.2 & -0.5$^{***}$ \\ 
Trend with control  & -0.2 & -0.1 & -0.3$^{*}$ & -0.2 & -0.1 & -0.5$^{**}$ \\ 
Trend with control (beta regression) & -1.9 & -1.7 & -3.8 & -2.0 & -1.8 & -5.7$^{**}$ \\ 
\end{tabular}}
\label{tab:ratios}
\end{table}

\newpage

\end{document}